\documentclass[fleqn,11pt]{article}
\usepackage{amsmath,amssymb,amsfonts,latexsym}
\usepackage{graphicx}

\def\eqn#1{\eq\eqref{#1}}
\def\rf{\eqref}

\def\MN{^{\mu\nu}}
\def\mN{_\mu^\nu}

\def\R{{\mathbb R}}

\def\tT{{\widetilde T}}

\def\kappa{\varkappa}

\def\grav{gravitational}

\def\cy{cylindrical}
\def\cyl{cylindrically symmetric}

\def\wh{wormhole}
\def\whs{wormholes}
\def\asflat{asymptotically flat}
\def\intcon{integration constant}

\usepackage{amsfonts,amssymb,cite}
\usepackage{graphicx}

\def\noi{\noindent}

\newcommand{\Title}[1]{\noi {{\Large\bf #1}}\\[1ex]}

\def\Aunames#1{\noi{\bf #1}}
\def\au#1{${}^{#1}$}
\def\Addresses#1{\medskip\noi \protect
	\begin{description}\itemsep -3pt {\it #1} \end{description}}
\def\adr#1#2{\item[${}^{#1}$]{\it #2}}

\newcommand{\Abstract}[1]{\vskip 2mm \begin{center}
        \parbox{16.4cm}{\small\noi #1} \end{center}\medskip}

\def\email#1#2{\footnotetext[#1]{e-mail: #2}\addtocounter{footnote}{1}}

\def\nq{\hspace*{-1em}}
\def\nqq{\hspace*{-2em}}
\def\nhq{\hspace*{-0.5em}}

\def\cm{\hspace*{1cm}}
\def\inch{\hspace*{1in}}

\def\Jl#1#2{#1 {\bf #2},\ }

\def\ApJ#1 {\Jl{Astroph. J.}{#1}}
\def\CQG#1 {\Jl{Class. Quantum Grav.}{#1}}
\def\DAN#1 {\Jl{Dokl. AN SSSR}{#1}}
\def\GC#1 {\Jl{Grav. Cosmol.}{#1}}
\def\GRG#1 {\Jl{Gen. Rel. Grav.}{#1}}
\def\IJMPD#1 {\Jl{Int. J. Mod. Phys. D}{#1}}
\def\JETF#1 {\Jl{Zh. Eksp. Teor. Fiz.}{#1}}
\def\JETP#1 {\Jl{Sov. Phys. JETP}{#1}}
\def\JHEP#1 {\Jl{JHEP}{#1}}
\def\JMP#1 {\Jl{J. Math. Phys.}{#1}}
\def\NPB#1 {\Jl{Nucl. Phys. B}{#1}}
\def\NP#1 {\Jl{Nucl. Phys.}{#1}}
\def\PLA#1 {\Jl{Phys. Lett. A}{#1}}
\def\PLB#1 {\Jl{Phys. Lett. B}{#1}}
\def\PRD#1 {\Jl{Phys. Rev. D}{#1}}
\def\PRL#1 {\Jl{Phys. Rev. Lett.}{#1}}

\def\al{&\nhq}
\def\lal{&&\nqq {}}
\def\eq{Eq.\,}
\def\eqs{Eqs.\,}
\def\beq{\begin{equation}}
\def\eeq{\end{equation}}
\def\bear{\begin{eqnarray}}
\def\bearr{\begin{eqnarray} \lal}
\def\ear{\end{eqnarray}}
\def\earn{\nonumber \end{eqnarray}}

\def\nnv{\nonumber\\[5pt] {}}
\def\nnn{\nonumber\\ \lal }
\def\nnnv{\nonumber\\[5pt] \lal }
\def\yy{\\[5pt] {}}
\def\yyy{\\[5pt] \lal }
\def\eql{\al =\al}

\def\tst{\textstyle}

\def\fract#1#2{{\tst\frac{#1}{#2}}}

\def\half{{\fract{1}{2}}}

\def\e{{\,\rm e}}

\def\sign{\mathop{\rm sign}\nolimits}
\def\diag{\mathop{\rm diag}\nolimits}

\def\const{{\rm const}}

\def\then{\ \Rightarrow\ }

\newcommand{\vars}[1]{\left\{\begin{array}{ll}#1\end{array}\right.}

\begin{document}
\thispagestyle{empty}
\twocolumn[

\Title{Rotating cylinders with anisotropic fluids in general relativity}

\Aunames{S. V. Bolokhov\au{a,1}, K. A. Bronnikov\au{a,b,c,2}, 
		and M. V. Skvortsova\au{a,3}} 

\Addresses{
\adr a {\small Peoples' Friendship University of Russia (RUDN University), 
             ul. Miklukho-Maklaya 6, Moscow 117198, Russia}
\adr b {\small Center for Gravitation and Fundamental Metrology, VNIIMS,
             Ozyornaya ul. 46, Moscow 119361, Russia}
\adr c {\small National Research Nuclear University ``MEPhI''
                    (Moscow Engineering Physics Institute), Moscow, Russia}
	}		
	
\Abstract
	{We consider anisotropic fluids with directional pressures $p_i = w_i \rho$ ($\rho$ is the density, 
	$w_i = \const$, $i = 1,2,3$) as sources of gravity in stationary cylindrically symmetric space-times. 	
	We describe a general way of obtaining exact solutions with such sources, where the main features 
	are splitting of the Ricci tensor into static and rotational parts and using the harmonic radial 
	coordinate. Depending on the values of $w_i$, it appears possible to obtain general or special
	solutions to the Einstein equations, thus recovering some known solutions and finding new ones.	
	Three particular examples of exact solutions are briefly described: with a stiff isotropic perfect 
	fluid ($p = \rho$), with a distribution of cosmic strings of azimuthal direction (i.e., forming circles 
	around the $z$ axis), and with a stationary combination of two opposite radiation flows 
	along the $z$ axis. 	
	} 
	
] 
\email 1 {boloh@rambler.ru}
\email 2 {kb20@yandex.ru}
\email 3 {milenas577@mail.ru}

\section{Introduction}

  Studies of stationary \cyl\ configurations in general relativity (GR) have a long history, beginning with  
  the Lanczos (1924) --- Lewis (1932) vacuum solution \cite{lanczos, lewis} and continuing till now. 
  Among motivations of such studies one can mention the relative simplicity of the \grav\ field 
  equations as compared to more realistic axial symmetry (to say nothing on the general 
  nonsymmetric case) and the possible existence of linearly extended structures like cosmic strings
  in the Universe. In addition, \cy\ symmetry is the simplest one that admits rotation and is most 
  suitable for studying rotational phenomena in GR, so not too surprising is the wealth of existing 
  stationary (that is, assuming rotation) exact solutions to the Einstein equations with various sources of
  gravity: the cosmological constant $\Lambda$ 
  \cite{vac-Lam1, vac-Lam2, vac-Lam3};
  scalar fields with or without self-interaction potentials \cite{BLem13,BK15}, including $\Lambda$ as 
  a constant potential \cite{erices}; dust in rigid or differential rotation 
  \cite{van_stockum, bonnor09, iva02c}, electrically charged dust \cite{iva02b}, 
  dust with a scalar field \cite{santos82}, perfect fluids with various equations of state, mostly 
  $p=w\rho$, $w=\const$ (in usual notations) \cite{hoen79, davids97, davids00, skla99, iva02a}; 
  some examples of anisotropic fluids \cite{anis1, anis2, anis3} etc, see also references therein
  as well as reviews  \cite{exact-book, BSW19}. 
  
  There are also many solutions to the field equations in various extensions of GR most 
  of which come into play at extremely large curvatures or matter densities, at very small or 
  very large length scales etc. If our interest is in macroscopic scales, say, from meters to 
  kiloparsecs, then, in our opinion, it makes sense to adhere to GR which is the simplest and 
  has a brilliant experimental status. Moreover, many solutions of alternative theories can be 
  obtained from those of GR by different solution generating methods, such as the well-known 
  conformal mapping connecting the Jordan and Einstein frames in scalar-tensor and $f(R)$  
  theories. 

  In this paper we undertake a systematic study of the Einstein equations for stationary \cyl\ 
  space-times where the source of gravity is an anisotropic fluid with each principal pressure   
  $p_i$ equal to density times a constant proportionality factor $w_i$. This ansatz includes as
  special cases perfect isotropic fluids with $p/\rho = w = \const$ and many anisotropic 
  sources of interest such as cosmic string bundles with radial, longitudinal or azimuthal directions,
  some stationary cases with radiation flows of the same directions, some field configurations, etc. 
  Our purpose here is to present a general method of solving the Einstein equations using the 
  fact that in stationary \cyl\ space-times the Ricci and Einstein tensors split into static and 
  rotational parts \cite{BLem13} and employing the harmonic radial coordinate, first applied
  to static \cy\ systems in \cite{kb79}. This leads to reproducing some known solutions and 
  obtaining new ones which can be applied to describe configurations with a regular symmetry
  axis and those where such an axis is absent (\cy\ \whs\ 
  \cite{BLem09, BLem13, BK15, BK18, BBS19}.  A thorough analysis of numerous 
  systems of interest is postponed for the future, here we only briefly consider some examples.   

\section{Stationary \cy\ space-\\ times}

\subsection{Basic relations}

  Consider a stationary \cyl\ metric
\bearr                                                    \label{ds-rot}
         ds^2 = \e^{2\gamma(x)}[ dt - E(x)\e^{-2\gamma(x)}\, d\varphi ]^2- \e^{2\alpha(x)}dx^2 
\nnn \cm
	- \e^{2\mu(x)}dz^2 - \e^{2\beta(x)}d\varphi^2,
\ear
  where $x$, $z\in \R$ and $\varphi\in [0, 2\pi)$ are the radial, longitudinal and 
  angular coordinates. The coordinate $x$ is specified up to a reparametrization
  $x \to f(x)$, so its range depends on both its choice (the ``gauge'') and the
  geometry itself. The off-diagonal component $E$  describes rotation, 
  and the corresponding vortex gravitational field in space-times with the 
  metric \rf{ds-rot}  is characterized by the angular velocity $\omega(x)$ 
  of a congruence of timelike curves (vorticity) \cite{BLem13, kr2, kr4},
\beq                                                  \label{om}
          \omega = \half (E\e^{-2\gamma})' \e^{\gamma-\beta-\alpha}.
\eeq
  under an arbitrary choice of the coordinate $x$ (a prime stands for
  $d/dx$). Furthermore, in the reference frame comoving to matter 
  in its motion in $\varphi$ direction, we have the SET component
  $T^3_0 = 0$, hence (via the Einstein equations) we have the Ricci tensor 
  component $R_0^3 \sim (\omega \e^{2\gamma+\mu})' = 0$, so that 
\beq       	      					\label{omega}
	\omega = \omega_0 \e^{-\mu-2\gamma}, \cm \omega_0 = \const.
\eeq
  Then, according to \rf{om}, 
\beq                          \label{E}
	E(x) = 2\omega_0 \e^{2\gamma(x)} \int \e^{\alpha+\beta-\mu-3\gamma}dx.
\eeq
  The nonzero components of the Ricci $(R\mN)$ are
 \bear                 \label{Ric}
      R^0_0 \eql -\e^{-2\alpha}[\gamma'' + \gamma'(\sigma' -\alpha')] - 2\omega^2,
\nnv      
      R^1_1 \eql -\e^{-2\alpha}[\sigma'' + \sigma'{}^2 - 2U - \alpha'\sigma']+ 2\omega^2,
\nnv      
      R^2_2 \eql -\e^{-2\alpha}[\mu'' + \mu'(\sigma' -\alpha')],  
\nnv      
      R^3_3 \eql -\e^{-2\alpha}[\beta'' + \beta'(\sigma' -\alpha')] + 2\omega^2, 
\nnv
      R^0_3 \eql G^0_3 =  E \e^{-2\gamma}(R^3_3 - R^0_0), 
\ear
 where we have introduced the notations
\beq
            \sigma = \beta + \gamma + \mu, \qquad
            U = \beta'\gamma'  + \beta'\mu' + \gamma' \mu'.
\eeq     
  The Einstein equations may be used in two equivalent forms 
\bearr    \label{EE1}
            G\mN \equiv R\mN - \half \delta\mN R = -\kappa T\mN, \quad {\rm or}
\\   \lal                 \label{EE2}
            R\mN = - \kappa \tT\mN \equiv -\kappa(T\mN - \half \delta\mN T).          
\ear    
  where $\kappa = 8\pi G$ is the gravitational constant, $R$ the Ricci scalar, and $T$ 
  the trace of the SET. In what follows we will mostly use the equations in the form \rf{EE2}, but 
  it is also helpful to join the constraint equation from \rf{EE1}, which is the first integral of the 
  others and contains only first-order derivatives of the metric functions:
\beq                  \label{G11}
	      G^1_1 = \e^{-2\alpha} U + \omega^2 = - \kappa T^1_1.
\eeq    
  
  One can see from \rf{Ric} that the diagonal components of the Ricci
  ($R\mN$) and Einstein ($G\mN$) tensors split into those for the static metric 
  (that is, the metric (\ref{ds-rot}) with $E=0$) plus a contribution from $\omega$
  \cite{BLem13}: 
\bearr 		\label{R-omega}
    		R\mN = {}_s R\mN + {}_\omega R\mN, \quad
	{}_\omega R\mN = \omega^2 \diag (-2, 2, 0, 2),  
\nnn	
\yyy     		\label{G-omega}  
		G\mN = {}_s G\mN + {}_\omega G\mN,  \quad
   	{}_\omega G\mN = \omega^2 \diag (-3, 1, -1, 1),  
\nnn
\ear
  where ${}_s R\mN$ and ${}_s G\mN$ are the static parts. The tensors ${}_s G\mN$ 
  and ${}_\omega G\mN$ (each separately)  satisfy the conservation law 
  $\nabla_\alpha G^\alpha_\mu =0$ according to this static metric. Thus the tensor 
  ${}_\omega G\mN/\kappa$ acts as an  additional SET with exotic properties 
  (e.g., the effective energy density is $ -3\omega^2/\kappa <  0$), favorable for 
  obtaining wormhole solutions, as confirmed by a number of examples in 
  Refs \cite{BLem13,BK15,kr4,BBS19}.
  
  Notably, it is sufficient to solve the diagonal components of the Einstein equations, 
  their single off-diagonal component then holds as well \cite{BLem13}.  

\subsection {Boundary conditions}

  {\bf 1. A regular symmetry axis.}
  In addition to the field equations, physical problems generally require imposing some
  well-motivated boundary conditions. In \cy\ symmetry, it is most often required that
  the space-time metric should be regular on the symmetry axis, which guarantees that 
  there is no source of gravity (say, a cosmic string along the axis) other than the one under
  consideration. A regular axis certainly assumes that at the corresponding value of the 
  radial coordinate, where $g_{\phi\phi} \to 0$, all curvature invariants tend to finite 
  limits, which, in terms of the metric  \rf{ds-rot} requires, in particular, finite limits of 
  $\beta, \gamma, \mu, E$. One more requirement (similar to spherical symmetry) is 
  that of a correct circumference to radius ratio ($2\pi$) for small circles around the axis.
  This is achieved as long as (see, e,g., \cite{anis3})\footnote
  	{Note that this condition, as all other formulas in this section,  is written in a form 
  	independent from the choice of the radial coordinate $x$.}
\beq             \label{axis}
                \frac 1{4X} \e^{-2\alpha} X'{}^2  \to 1,    \qquad  X = |g_{\varphi\varphi}|.
\eeq      
  If $E =0$ on the axis, this condition reduces to $\e^{-2\alpha + 2\beta}\beta'{}^2 \to 1$
  (quite similar to spherical symmetry where $\e^{2\alpha} = -g_{11}$ and $\e^{2\beta}$ 
  is the coefficient near the angular part of the metric \cite{ws-book}).  
  It prevents the occurrence of a conical singularity on the axis and is necessary for the 
  existence of a unique tangent flat space at points located on the axis.
  
\medskip\noi  
  {\bf 2. The outer boundary of a fluid distribution.}  
  A particular solution to the field equations may describe a fluid distribution that occupies the 
  whole space, but in many cases it proves to be necessary to restrict this distribution to
  a finite region outside which space-time is empty or filled with another kind of matter, such 
  as an electromagnetic or scalar field, a cosmological constant, etc. At a boundary surface 
  $\Sigma$ between two media, it is required that the metric tensor should be continuous 
  (otherwise it is impossible to say that this boundary as seen ``from the left'' and ``from the 
  right'' is the same surface). In addition (see, e.g., \cite{darmois, israel67, BKT87}), the extrinsic 
  curvature tensor $K_{ab}$ of $\Sigma$ (the indices $a,b$ correspond to coordinates on $\Sigma$) 
  must be continuous across it. For \cyl\ space-times, in  which case $\Sigma$ is a cylinder, 
  continuity of $K_{ab}$ implies continuity of the radial pressure $p_r$, hence, in particular, 
  matching to a vacuum region with $T\mN =0$ is only possible on a surface on which $p_r =0$.
  
  If we admit a finite discontinuity of $K_{ab}$, it means that we admit $\Sigma$ as a thin 
  shell with a surface SET proportional to this discontinuity.  

\medskip\noi                    
{\bf 3. Wormhole geometries.} It can happen that a \cyl\ space-time does not contain a symmetry
  axis, but, instead, the circular radius $r(x) = \e^{\beta(x)}$ has a regular minimum on a certain
  cylinder $x = x_0$, called a throat\footnote
	{There can be another definition of a \cy\ throat, related to the behavior of the area 
	function $a(x)=\e^{\beta + \mu}$ instead of $\e^\beta$, see a discussion in 
	\cite{BLem09, BLem13}.}         
  and is large or infinite far from this minimum. Then it makes sense to discuss the boundary 
  conditions at such large radii or outer boundaries on both sides of the throat.

\medskip\noi    
{\bf 4. Asymptotic flatness.}  If an object under consideration is to be observable from
  a distant, almost flat space-time region, the corresponding metric should be \asflat.
  A shortcoming of almost all \cyl\ solutions is a lack of asymptotic flatness: even the 
  simplest static vacuum Levi-Civita solution is \asflat\ only if it is completely flat.
  This problem is ignored in many studies, in which static \cy\ solutions are matched on
  some surfaces with a Levi-Civita exterior, and stationary ones with a Lewis exterior. 
  This problem was discussed for \wh\ models in 
  \cite{BLem13,BK15,kb-conf16,BK18}, where, following the suggestion of 
  \cite{BLem13}, \wh\ space-times were constructed with three regions: a central 
  one containing a throat and two external Minkowski regions matched to this central one
  at some cylinders $\Sigma_-$ and $\Sigma_+$. These surfaces inevitably 
  contained some densities and pressures calculated from jumps of $K_{ab}$ across 
  them. A separate problem is to find such models where matter in the central region and 
  on its both boundaries satisfies the Weak and Null Energy Conditions. These issues 
  are  beyond the scope of the present paper, for details see 
  \cite{BLem13,BK15,kb-conf16,BK18,BBS19}.
  
\section {Solutions with anisotropic fluids}

\subsection{Anisotropic fluids: the formalism} 

  The general formalism of nondissipative anisotropic fluids in Riemannian space-time 
  \cite{anis1,anis2,anis3}
  assumes matter with a certain energy density $\rho$ and three principal pressures $p_i$ in 
  mutually orthogonal directions, so that the SET has the form\footnote
  	{Tetrad indices are written in parentheses.} 
\beq              \label{SET1}
			T^{(\alpha\beta)} = \diag (\rho, p_1, p_2, p_3)
\eeq
  in an orthonormal tetrad corresponding to the fluid's comoving reference frame (its comoving 
  nature is evident from the zero values of the energy flux components $T^{(0i)}$, $i =1,2,3$) 
\beq                 \label{tet}
                       \big( e^\mu_{(\alpha)} \big) = \big( u^\mu, \phi^\mu, \chi^\mu, \psi^\mu \big), 
\eeq    
  where $u^\mu$ is timelike and has the meaning of the fluid's 4-velocity while the other tetrad vectors 
  $\phi^\mu, \chi^\mu, \psi^\mu$ are spacelike. The usual coordinate SET components are 
\bearr 	             \label{SET2}
  	T\MN = e^\mu_{(\alpha)}e^\nu_{(\beta)} T^{(\alpha\beta)} 
\nnn \ \  	
  	  = \rho u^\mu u^\nu  + p_1\phi^\mu \phi^\nu + p_2\chi^\mu \chi^\nu + p_3\psi^\mu \psi^\nu. 
\ear  			   
  A more conventional form of $T\MN$ is obtained from \rf{SET2} by adding and subtracting the 
  quantity $p_1 g\MN$, where the choice of $p_1$ among the pressures is arbitrary, and taking 
  into account that 
\[    \nhq
  g\MN =  e^\mu_{(\alpha)}e^\nu_{(\beta)} \eta^{(\alpha\beta)} 
  =  u^\mu u^\nu  - \phi^\mu \phi^\nu - \chi^\mu \chi^\nu - \psi^\mu \psi^\nu.
\]
  We thus obtain  
\bearr                         \label{SET3}
	T\MN = (\rho + p_1) u^\mu u^\nu  - p_1 g\MN 
			+ (p_2 - p_1)\chi^\mu \chi^\nu 
\nnn \cm			
			+ (p_3 - p_1)\psi^\mu \psi^\nu. 
\ear    
  If $p_1 =p_2 =p_3 =p$, we obtain the usual (isotropic) perfect fluid SET, 
  $T\MN = (\rho+p) u^\mu u^\nu - p g\MN$. 

\medskip
  Let us now consider such matter in space-times with the metric \rf{ds-rot}, assuming that $p_1= p_x$ 
  corresponds to the radial pressure, $p_2 = p_z$ and $p_3 = p_\varphi$ to pressures in the $z$ and 
  $\varphi$ directions. Then, as is easily verified,  the conservation law  $\nabla_\nu T\mN =0$ takes the 
  explicit form:
\beq 				\label{cons}
		p'_x + (\rho + p_x) \gamma' + (p_x - p_z) \mu' + (p_x - p_\varphi) \beta' =0. 
\eeq    
  It is of interest that $-g_{03} = E$ does not appear in \rf{cons}, and this relation is the same as with
  the static metric. 

\subsection{A search for solutions}

  Consider the simplest class of equations of state for the SET \rf{SET3}:
\beq
	   p_x = w_1 \rho, \qquad p_z = w_2 \rho, \qquad p_\varphi = w_3 \rho,
\eeq    
  with $w_i = \const$. Up to now, we worked with an arbitrary radial coordinate $x$. To solve the Einstein 
  equations \rf{EE2}, it is helpful to choose the harmonic ``gauge'' 
\beq                                                   \label{harm}
		\alpha = \beta + \gamma + \mu.
\eeq     
    Then the components ${0\choose 0}, {2\choose 2}, {3\choose 3}$  of \rf {EE2} and \eqn{G11} 
    can be written in the form 
\bearr         			\label{EE00}
		\e^{-2\alpha} \gamma'' + 2 \omega^2 = \frac{\kappa\rho}{2} (1 + w_1 + w_2 + w_3),
\yyy        \qquad        \ \  			\label{EE22}
		\e^{-2\alpha} \mu''  = \frac{\kappa\rho}{2} (-1 + w_1 - w_2 + w_3),
\yyy	          			\label{EE33}								
		\e^{-2\alpha} \beta'' - 2 \omega^2 = \frac{\kappa\rho}{2} (- 1 + w_1 + w_2 - w_3),
\yyy                   \label{int}
	\e^{-2\alpha}(\beta'\gamma'\! + \beta'\mu' \! + \gamma' \mu')\! + \omega^2 	= w_1\kappa \rho,  
\ear            
   respectively. The sum of \rf{EE00} and \rf{EE33} gives
\beq                                               			\label{EE03}
			\beta'' + \gamma'' = \kappa\rho \e^{2\alpha} (w_1 + w_2).
\eeq   		   		
   Equation \rf{EE22} also does not contain $\omega$, and combining it with \rf{EE03}, we obtain 
   an easily integrable relation involving only second-order derivatives of the metric functions
\beq                                                  \label{ww}
			2(w_1 + w_2) \mu'' = (-1 + w_1 -w_2 + w_3) (\beta'' + \gamma'').  
\eeq 
   This relation enables us to exclude one of the three unknowns $\beta, \gamma, \mu$ from our equations. 
   
    Next, we note that, if we leave $\beta'', \gamma'', \mu''$ in any linear combination of \eqs 
   \rf{EE00}--\rf{EE33}, the r.h.s. will contain a linear combination of two terms, one proportional to 
\beq                                                    \label{def-xi}
			\omega^2 \e^{2\alpha} = \omega_0^2 \e^{2\xi}, \qquad \xi := \beta - \gamma,  
\eeq            
   and the other proportional to $\rho \e^{2\alpha}$. Assuming $w_1 \ne 0$, the conservation equation 
   \rf{cons} is integrated giving 
\bearr                                                              \label{rho-gen}
	\rho = \rho_0 \exp\bigg\{\! - \frac 1 {w_1} \Big[(1+w_1)\gamma + (w_1 - w_2)\mu
\nnn \cm
			 + (w_1 - w_3)\beta\Big]\bigg\}, \qquad  \rho_0 = \const.
\ear
   Therefore, recalling \rf{cons}, we can write       
\bearr                                                   \label{def-eta}
			\rho \e^{2\alpha} = \rho_0 \e^{2\eta}, 
\nnnv  
			2\eta:= \beta \Big(1\!+\! \frac {w_3}{w_1}\Big) + \gamma\Big(1\!- \!\frac 1{w_1}\Big)
				+\mu \Big(1 \! + \! \frac {w_2}{w_1}\Big).
\ear     
  Taking into account the relationship \rf{ww} and combining the equations \rf{EE00}--\rf{EE33}, 
  it is straightforward to obtain coupled equations for $\xi$ and $\eta$:
\bear                 \label{xi''} 
			\xi''  \eql A \e^{2\xi} + B \e^{2\eta},
\yy                                 \label{eta''}
			\eta'' \eql C \e^{2\xi} + D \e^{2\eta},
\ear
   where the constant factors $A,B,C,D$ depend on the values of $w_i$ as well as $\rho_0$ and $\omega_0$.
   If $B = C = 0$, we have two separate Liouville-type equations which are solved directly and completely,
   and it remains to substitute the solution to the first-order equation \rf{int} to verify the solution and to 
   obtain a relation between the integration constants.
   
   Apart from the case $B = C = 0$, in which the matrix of coefficients in \eqs \rf{xi''}, \rf{eta''} is diagonal,
   this set of equations can be integrated if this matrix is degenerate, that is, its determinant 
   $AD - BC$ is zero. Indeed, in this case we have 
\beq
			A\eta'' = C \xi'' \ \ \Leftrightarrow \ \ B\eta'' = D \xi'',
\eeq                    
   so that $\eta = (C/A) \xi + \eta_1 x + \eta_0$, where $\eta_1, \eta_0 = \const$, and at least in the case 
   $\eta_1 =0$ we then obtain for $\xi(x)$ an equation integrable by quadratures
\beq                                  \label{xi''deg}
		 \xi'' = A \e^{2\xi} + B \e^{2(C/A)\xi + 2\eta_0}.
\eeq      
   A nonzero $\eta_1$ with explicit $x$-dependence may prevent such easy integration. Thus, in general,
   in this case we can obtain an ``almost general'' solution with one lacking integration constant.  
   There exist other completely integrable cases related to Toda systems, see, e.g., \cite{ivas02}.

   If the matrix of coefficients in \eqs \rf{xi''}, \rf{eta''} is neither diagonal nor degenerate, it is still 
   often possible to obtain special solutions of interest, as is described below. The case $w_1 =0$
   should be considered separately.
     
\subsection{A general solution}

  Consider the general case $w_1 \ne 0$ and assume $w_1+w_2 \ne 0$, then \eqn{ww} allows us to 
  express $\mu$ in terms of $\beta$ and $\gamma$. Indeed, integration of \rf{ww} gives
\beq          \label{mu}
	\mu = \frac {-1+w_1-w_2+w_3}{2(w_1+w_2)}\big( \beta+\gamma + mx \big),
\eeq      
  where $m=\const$ and another \intcon\ is suppressed by choosing a scale along the $z$ axis.
  Then, after some algebra, we obtain the following coefficients in \eqs \rf{xi''} and \rf{eta''}:
\bearr                            \label{ABCD}
		A = 4\omega_0^2,
\nnn
		B = -\kappa\rho_0 (1+w_3),
\nnn
		C = \frac{\omega_0^2}{w_1} (1+w_3),
\nnn
		D = \frac {\kappa \rho_0}{4w_1} \big[(w_1+w_2)(3w_1-w_2+2w_3-2) 
\nnn \inch		
		- (1+w_3)^2\big],
\ear    
  while the unknown functions $\xi(x)$ and  $\eta(x)$ are
\bearr
		\xi = \beta - \gamma,
\nnn
		\eta = (3w_1-w_2+3w_3-1)\frac{\beta}{4w_1} 
\nnn \cm		
		+ (3w_1-w_2+w_3-3)\frac{\gamma}{4w_1} + Kmx,
\nnn \
	   {\rm with} \quad	K := \frac 14 (w_1-w_2+w_3-1).				
\ear    
  The set of equations \rf{xi''}, \rf{eta''} can be completely integrated in the case $w_3 = -1$.
  That is, assuming $w_1 \ne 0$, $w_1 + w_2 \ne 0$, $w_3 = -1$, we have the following 
  decoupled equations  for $\xi(x)$ and $\eta(x)$:
\beq               \label{decoup}
		\xi'' = 4\omega_0^2 \e^{2\xi}, \cm \eta'' =  D \e^{2\eta}
\eeq  
 with $D =  \kappa \rho_0 (w_1+w_2)(3w_1 - w_2 - 4)/(4 w_1)$. These equations are easily solved, 
 and the specific form of the solutions will depend on the sign of $D$. We leave their full analysis for 
 the future. 
 
 The other integrable case, $AD = BC$, takes place if $3 w_1 -w_2 + 2w_3 = 2$, but the resulting 
 equation \rf{xi''deg} for $\xi(x)$ does not in general lead to solutions in elementary functions. 
 
\subsection{Special solutions for $w_1 \ne 0$}

 If $w_3 + 1 \ne 0$, it is in general impossible to solve the coupled equations \rf{xi''} and \rf{eta''}.completely,
 but some special solutions can be obtained. Indeed, let us suppose  
\beq             \label{eta-xi}
              \eta = \xi + \ln h, \cm h = \const,                          
\eeq  
 so that $\xi'' = \eta''$, and, comparing \eqs \rf{xi''} and \rf{eta''}, we obtain the consistency requirement . 
\bearr                \label{ABh}
              A + Bh^2  = C + D h^2, 
\nnn 
	{\rm or}\qquad  A-C = (D-B) h^2.     
\ear
  Note that $A -C$ is proportional to $\omega_0^2$ and $D-B$ to $\kappa \rho_0$, but these two parameters 
  are still not directly related due to the existence of the arbitrary factor $h^2$. Evidently, a solution only exists 
  if and only if the signs of $A-C$ and $D-B$ are the same. If it is the case, we have
\beq               \label{xi''a}
              \xi'' = (A + Bh^2) \e^{2\xi},
\eeq
  again a Liouville equation which is easily solved. After that, knowing $\xi = \beta - \gamma$ and a relation between 
  $\beta$ and $\gamma$ that follows from \rf{eta-xi}, one easily finds $\beta(x)$ and $\gamma(x)$ and then
  $\mu(x)$ from  \rf{mu}. To know the metric completely, it remains to substitute the results for $\beta, \gamma, \mu$ 
  to \rf{G11} and \rf{E} to find a relation between integration constants and to determine $E(x)$.

\subsection{Fluids with $w_1 \ne 0$, special cases of \eqn{ww}}

  {\bf 1. Fluids with  $w_1 + w_2 = 0$ but $1+w_2 \ne w_1 + w_3$.}   
  We return to the set of equations \rf{EE00}--\rf{EE33}, and \rf{ww} as their consequence, but the 
  l.h.s. of the latter is now zero, and the r.h.s then yields 
\beq              \label{bg}
	\beta'' + \gamma'' =0 \ \ \then \ \ \beta = -\gamma + b x + \ln r_0,
\eeq
  with $b, r_0 = \const$, and both $\xi$ and $(eta$ are expressed in terms of $\gamma$:
\bearr
		\xi = -2\gamma + bx + \ln r_0, 
\nnn    
	       \eta = \frac {-(1+w_3)\gamma + (w_1+w_3)(bx + \ln r_0)}{2 w_1}.
\ear     
  Therefore, \eq  \rf{EE00} contains only one unknown function $\gamma(x)$, though, it is uneasy to solve it
  in the general case $b\ne 0$. Having solved it, it is then easy to find $\mu(x)$ by directly integrating 
  \eqn {EE22}.
  
  In the special case $b=0$, it makes sense to use the constraint \rf{int}, which is then a first integral of 
  \rf{EE00}: we have $\beta'+\gamma' =0$, so that $\mu'$ is excluded from \rf{int} that now reads
\beq    \nq\,
		\gamma'{}^2 = \omega_0^2 r_0^2 \e^{-4\gamma} 
				- \kappa \rho_0 w_1 r_0^{1 {+} w_3/w_1} \e^{-(1{+}w_3)\gamma/w_1}
\eeq       
  and can be solved by quadratures (though in known functions only for some values of $w_1$ and $w_3$).
  And, as in other cases, already knowing $\beta, \gamma, mu$, it remains to find $E(x)$ from \rf{E}.        

\medskip\noi
{\bf 2. Fluids with  $1+w_2 = w_1 + w_3$ but $w_1 + w_2 \ne 0$.}    
 Under these assumptions, instead of \rf{bg}, we obtain from \rf{ww} (or equivalently from \rf{EE22})
\bearr       \label{mu-c}      
		\mu'' =0 \ \ \then \ \ \mu = mx + m_0,  
\nnn 
		m, m_0 = \const,
\ear
 where we can get $m_0=0$ by rescaling the $z$ axis. 
 
 It turns out that in this case the situation with solving our set of equations is practically the same as in the 
 general case. Thus, we have \eqs \rf{xi''} and \rf{eta''} that decouple under the condition $w_3 = -1$,
 in which case we have $B = C =0$ and arrive at \eqn{decoup} with $D = \kappa \rho_0 (w_1-1)^2 /w_1$.
 
 If $w_3 \ne -1$, some special solutions can be obtained in the same manner as with \eqs \rf{eta-xi}--\rf{xi''a}.
    
\medskip \noi
{\bf 3. Fluids with $w_1 - w_2 + w_3 -1 = w_1 + w_2 = 0$.}
 Now both equations \rf{bg} and \rf{mu-c} are valid, and there remains only one unknown function $\gamma(x)$
 in the set   \rf{EE00}--\rf{EE33}. Moreover, for $w_i$ we have
\beq
		w_2 = -w_1, \qquad w_3 = 1 - 2w_1,
\eeq        
  and the functions $\xi(x)$ and $\eta(x)$ are expressed as
\bearr
		\xi = - 2\gamma + bx + \ln r_0, 
\nnn 
		\eta = \frac {1-w_1}{2w_1}( - 2\gamma + bx + \ln r_0).
\ear                      
  Also, as in the previous case, we have $\mu''=0  \then \mu = mx$.
  The function $\gamma(x)$ can be found by quadratures from the equation that follows from \rf{int},
\beq  
 		\gamma'{}^2 - b\gamma' - bm = \omega_0^2 \e^{2\xi} - \kappa\rho_0 w_1 \e^{2\eta}.
\eeq 
  Thus in this case the problem is solved completely by quadratures. 

\subsection{Solutions with $w_1 =0$}  
      
  We so far  assumed $w_1 \ne 0$. What changes if $w_1 =0$? In such a case 
  \eqn{cons} yields (provided $\rho \ne 0$) $\gamma' = w_2 \mu' + w_3 \beta'$, whence
\beq                 \label{ww1}  
		\gamma = w_2 \mu + w_3 (\beta - \ln r_0), .
\eeq    
  where $r_0 = \const$ has the meaning of a length scale because $\e^\beta$ has the dimension of 
  length.  As before, from \rf{EE00}--\rf{EE33} we have the relation \rf{ww} which now reduces to
\beq                                                  \label{ww0}
			2w_2 \mu'' = (-1  -w_2 + w_3) (\beta'' + \gamma'').  
\eeq   
  Taken together, \rf{ww0} (after integration) and \rf{ww1} leave only one unknown among 
  $\beta, \gamma, \mu$. For this remaining function, the corresponding equation should be 
  combined again from \rf{EE00}--\rf{EE33}, depending on the values of $w_2$ and $w_3$,
  after which it will be straightforward to find $\rho(x)$, and, as before, the last steps in obtaining the 
  solution are a substitution to \rf{G11} leading to a relation between the integration constants, and 
  finding $E(x)$ by integration in \rf{E}.
  
  It should be noticed that under the condition $w_1 =0$ the anisotropic fluid distribution may be 
  in general cut at any value of $x$ and matched to the external vacuum (Lewis) solution without 
  any thin shells with nonzero surface stress-energy.

\subsection{Isotropic perfect fluids, $w\ne 0$}

  In the important case of $p = w\rho$, where $w = w_1=w_2=w_3$, the general solution available at $w_3=-1$   
  corresponds to a cosmological constant $\Lambda =  \rho/\kappa = -p/\kappa$. We thus arrive at the Lewis 
  well-known solution, which is traditionally written in other notations \cite{lanczos, lewis} and is also discussed in 
  \cite{exact-book,BSW19}; note also its generalizations with a massless scalar field 
  \cite{BLem13,erices} and scalar fields with exponential potentials \cite{BK15}.   
  
  At $w \ne -1$ we can hope to obtain special solutions as described in the previous subsection. Assuming 
  $\eta = \xi + \ln h$, we have 
\bearr
		\eta = \frac 1 {4 w}\big[ (5w-1)\beta + 3 (w-1)\gamma + (w-1) mx \big],
\nnnv
	         D = \frac {\kappa\rho} {4 w} (7 w^2 -6 w -1),  
\ear         
  and evident expressions for $A, B, C$ from \rf{ABCD}. Equation \rf{ABh} takes the form
\beq           \label{w-signs}
		4 \omega_0^2 (3w-1) = \kappa \rho_0 h^2 (11 w^2 - 2w -1).
\eeq    
  It can be easily verified that both sides of this equality are nonzero and have the same sign if 
\beq          \label{w-range}  \nhq
		w \in \bigg( \frac {1- 2\sqrt{3}}{11}, \frac 13 \bigg)  \ \	
		{\rm or} \ \  w >  \frac {1 +2\sqrt{3}}{11} \approx 0.406
\eeq      
 (both parts of \rf{w-signs} are negative in the first range and positive in the second one). Only in these ranges of 
 $w$ we can obtain solutions. Notably, two physically distinguished cases $w= 1/3$ (chaotic radiation, or an
 ultrarelativistic gas) and $w = -1/3$ (a chaotic gas of cosmic strings) do not belong to the range \rf{w-range}. 
 In particular, under the assumption $w=1/3$  \eqn{w-signs} leads to $\rho_0 = 0$, that is, the absence of a matter
 distribution. 
 
 With \rf{ABCD} and \rf{w-signs}, \eq \rf{xi''a} takes the Liouville form
\beq
		\xi'' = \frac{4w(2w -1)}{3w-1}  \kappa \rho_0 h^2 \e^{2\xi}.
\eeq  
 The sign of its r.h.s., determining the form of its solutions, is thus different for different $w$.  
 
 Solutions for arbitrary $w$ were obtained in \cite{iva02a} using a very unusual coordinate condition for $x$ 
 and reducing the problem to a special form of the Emden-Fowler equation. The presently described method 
 leads to more special solutions of a much simpler form.       
                
\section{Some special solutions}  
  
  We have described the way of obtaining exact solutions for quite diverse isotropic and 
  anisotropic fluid distributions. Their detailed description would be too long for this paper, 
  and we here restrict ourselves to some simple examples. 
  
\subsubsection*{Example 1: Stiff perfect fluid, $w=1$}      

  As is clear from Subsection 3.5, for perfect fluids with $w\ne -1$ our approach gives only special solutions
  and only in the ranges \rf{w-range} of the parameter $w$. Such a solution is especially simple for a 
  maximally stiff perfect fluid, $w=1$, for which the speed of sound is equal to the speed of light. Indeed, 
  \eqn{EE22} leads to $\mu'' =0$, hence we can write $\mu = mx$, $m = \const$. The assumption    
  \rf{eta-xi} then leads to   
\beq              \label{xi''w1}
                 \xi'' = 2 \omega_0^2 \e^{2\xi},      \qquad  \omega_0^2 = \kappa \rho_0 h^2,
\eeq    
  while since now we have $\eta = \beta + \mu = \beta + mx$ while $\xi = \beta-\gamma$, \eqn{eta-xi} 
  gives $\gamma = - mx + \ln h$. Furthermore, adjusting the time scale, we can assume $h =1$, so that
\beq                              \label{bgm-w1}
        	\mu = mx, \qquad \gamma = -mx, \qquad \beta = \xi + mx.
\eeq   
   Already at this stage we can substitute \rf{bgm-w1} to \eqn{G11}, where $\beta'$ is excluded due to 
  $\mu' + \gamma' =0$, and due to the second equality \rf{xi''w1} we obtain $ m=0$, so that we have 
  simply $\xi = \beta$. According to \rf{rho-gen}, in this configuration the density $\rho$ is constant.
  
  The Liouville equation \rf{xi''w1} for $\xi = \beta$ has the first integral 
\beq
		\beta'{}^2 = 2\omega_0^2 \e^{2\beta} + k^2 \sign k, \quad k = \const,
\eeq    
  and its solution splits into three branches according to the sign of $k$:
 \bearr                       \label{beta-w1}
 		\e^{\beta} = \frac{k_1}{\sqrt{2\omega_0^2} \cos (k_1 x)}, \quad  k = - k_1 <0, 	
\nnn
 		\e^{\beta} = \frac{1}{\sqrt{2\omega_0^2} x}, \qquad  k = 0,
\nnn
 		\e^{\beta} = \frac{k}{\sqrt{2\omega_0^2} \sinh (k x)}, \quad  k > 0. 		 		
\ear
  where in all cases an integration constant is excluded by choosing the zero point of $x$.
  The first branch corresponds to a metric of \wh\ nature, with $e^\beta$ having a minimum at $x=0$, and  
  $\beta \to \infty $  as $x \to \pm \pi/(2k_1)$. In this case \eqn{E} gives 
\beq   
  	E = \frac {k_1}{\omega_0} \tan(k_1 x) + E_0, \qquad E_0 = \const.
\eeq  	 
  This solution was used in \cite{BBS19} for obtaining a twice \asflat\ \cy\ \wh\ model by cutting it at sufficiently 
  small $|x| < \pi/(2k_1)$ and joining it to flat space regions on both sides from the throat $x=0$ through thin 
  shells.   
  
  The other two branches describe a full range of circular radii $r = \e^{\beta}$ from $r=0$ at $x = \infty$ 
  to $r \to \infty$ as $x \to 0$. For $E(x)$, \eqn{E} gives
\bearr
                 E = \frac 1 {\omega_0 x} \qquad (k=0), 
\nnn
	     E = \frac k {\omega_0} [1-\coth (kx)] \qquad (k >0), 
\ear  
  where we chose the integration constant in \rf{E} so that $E \to 0$ as $x \to \infty$.  
  It can be verified using \eqn{axis} that the axis is regular in the only case $k=1$ belonging to the third branch 
  \rf{beta-w1}. To our knowledge, this solution was obtained for the first time in other notations in \cite{hoen79}.  

\subsubsection*{Example 2: Azimuthal strings}

  A distribution of closed cosmic strings allocated in the $\phi$ direction corresponds to 
\beq                   \label{00-1}
  	 (w_1, w_2, w_3) = (0, 0, -1),
\eeq
  so that the tension $-p_\varphi$ in the angular $\varphi$ direction is equal to the energy density $\rho$.
  For this set of $w_i$, \eqn{ww1} is applicable, now leading to 
\beq  
         \beta = - \gamma + \ln r_0,  \qquad \xi = -2\gamma + \ln r_0,    
\eeq
  where, as before, $r_0$ is an arbitrary length scale. It turns out that in this case our set of field
  equations  is underdetermined since both \rf{EE00} and \rf{EE33} lead to the same equation 
  for $\gamma(x)$, 
\beq           \label{ga''}
		\gamma'' = -2\omega_0^2 r_0^2 \e^{-4\gamma},
\eeq    
  and \eqn{int} yields their integral,
\beq            \label{ga'}
  		\gamma'{}^2 = \omega_0^2 r_0^2 \e^{-4\gamma},
\eeq  
  whence it follows
\beq
		\e^{2\gamma} = 2r_0|\omega_0|x, \qquad \e^{2\beta} = \frac {r_0}{2|\omega_0|x}
\eeq    
  (as before, we choose the zero point of $x$ as to kill an insignificant integration constant). The 
  density $\rho$ dropped out from both \rf{ga''} and \rf{ga'}, and there is a single \eqn{EE22} 
  for the two unknowns $\mu$ and $\rho$:
\beq                  \label{muu}
		\mu'' = -\kappa r_0^2 \rho \e^{2\mu}.
\eeq    
  So one can take either $\mu(x)$ or $\rho \e^{2\mu}$ as an arbitrary function and calculate the
  remaining unknowns with \eqn{muu}. This means, in fact, that a set of rotating azimuthal cosmic 
  strings can have an arbitrary density distribution along the radial direction. 
  
  The last unknown $E$ is obtained from \rf{E} as
\beq
		E = r_0 (\sign \omega_0) (E_0 x - 1), \qquad E_0 = \const.
\eeq      
  Choosing $E_0 =0$, we obtain $E = -r_0 \sign \omega_0 = \const$ (which certainly does not 
  mean zero vorticity). 
  
  The solution is defined for $x \in \R_+$, $x \to 0$ corresponds to large radii $r = \e^\beta$ with 
  an attracting singularity due to $\e^\gamma \to 0$, while $x \to \infty$ leads to $r \to 0$ and large 
  positive $g_{33} = - \e^{2\beta} + E^2 \e^{-2\gamma}$, therefore circles parametrized by $\varphi$ 
  are closed timelike curves violating causality.  
  
\subsubsection*{Example 3: Longitudinal radiation flows}  

  One more situation of interest admitting a description in terms of anisotropic fluids is that there are 
  two opposite radiation flows of equal intensity in the $z$ direction, such that the net flux is zero. 
  
  If there are radiation flows with intensities $\Phi_+$ and $\Phi_-$ in the positive and negative $z$ 
  directions, respectively, the SET of each has the form $T\mN = \Phi_\pm k_\mu k^\nu$, where 
  $k^\mu$ is one of the the null vectors $(\e^{-\gamma}, 0, \pm \e^{-\mu}, 0)$. Then the nonzero part 
  of the summed SET of these two flows reads
\beq
              (T^a_b) = \begin{pmatrix} 
               				 \Phi_+ + \Phi_-& \Phi_+ - \Phi_- \\ 
               				 \Phi_+ - \Phi_- & -\Phi_+ - \Phi_-   
               				\end{pmatrix}, 
\eeq     			
  where $ a, b = 0, 2$.(A similar construction was discussed in \cite{we-kim16} in relation 
  to generalizations of the Birkhoff theorem.)  If $\Phi_+ = \Phi_- = \Phi$, the summed SET has the form 
\beq
		T\mN  = \rho \ \diag (1, 0, -1, 0), \qquad  \rho = 2\Phi. 
\eeq      
  It may be considered as the one of an anisotropic fluid with        
\beq                   \label{010}
  	 (w_1, w_2, w_3) = (0, 1, 0).
\eeq
    With such $w_i$, \eqs \rf{ww} and \rf{ww1} lead to 
\bearr  
         \beta''+ \gamma'' + \mu'' =0, 
\nnn         
           \gamma' = \mu'  \ \then \ \gamma = \mu      
\ear
  (with a proper choice of scales along the $z$ and $t$ axes), 
  and we can write, in the previous manner,
\beq              \label {b3}
	 \beta = - 2\gamma + bx + \ln r_0, \quad b, r_0 = \const. 
\eeq        
  Thus $\beta$ and $\mu$ are expressed in terms of $\gamma$, but it is reasonable to use as 
  the remaining unknown  
\beq              \label{x3}
		\xi(x) = \beta - \gamma = - 3\gamma + bx + \ln r_0,
\eeq    
  for which \eqn{int} leads to the easily integrable equation 
\beq                                                       \label{int3}
	\xi'{}^2 = 3\omega_0^2 \e^{2\xi} + b^2.
\eeq    
  Its solution is
\bearr
         \e^{-\xi} = \frac 1 {|b|} \sqrt{3\omega_0^2} \sinh (|b| x), \quad   b \ne 0,
\nnn
	 \e^{-\xi} = \sqrt{3\omega_0^2} x, \qquad b =0.
\ear	          
  and without loss of generality we assume $x >0$. Then for $b \ne 0$ the metric 
  coefficients are
\bearr
	\e^{3\gamma} = \e^{3\mu} = \frac{r_0}{2|b|} \sqrt{3\omega_0^2}\big|\e^{2bx}-1|,
\nnn
	\e^{3\beta} = 	 \frac{r_0}{b^2} \e^{bx} \sinh^2 (bx),
\nnn 
	E = (\sign \omega_0)(E_0 -  \coth(|b|x))
\nnn \inch \times	
			\bigg[\frac {2 r_0^2 |b|} {9 |\omega_0|}(\e^{2bx} - 1)^2 \bigg]^{1/3}.
\ear     
  In the case $b =0$ the metric coefficients are  
\bearr
	\e^{3\gamma} = \e^{3\mu} = \sqrt{3\omega_0^2} r_0 x,
\qquad
	\e^{3\beta} = \frac {r_0}{3\omega_0^2 x^2},
\nnn
	E = \frac {2\omega_0} {(3\omega_0^2)^{1/3}}  (r_0 x)^{2/3}  \Big(E_0 - \frac 1 x \Big).		
\ear 
  Lastly, the density can be determined from \eqn{EE22}: for any value of $b$:
  substituting $\alpha = \beta+\gamma+ \mu$ and \eqs \rf{b3} and \rf{x3}, we find
\beq  
		\kappa\rho r_0^2 = 2\omega_0^2 \e^{2\xi - 2bx}.
\eeq  
   and consequently
\beq
		\kappa\rho r_0^2 = \vars{   \dfrac{8b^2}{3(\e^{2bx}-1)^2},   \ &  b \ne 0,\\[10pt]
								2/ (3x^2)                             & b=0.
								} 
\eeq                        

  The solution is defined for $x \in \R_+$. We see that in all cases the value $x=0$, corresponding to 
  large radii $r = \e^\beta$, is a singularity with $\rho \to \infty$. However, other properties of the
  solution strongly depend on the sign of $b$ and, to a smaller extent, on the value of $E_0$.
  For example, as $x \to \infty$ (small $r$), $\rho \to 0$ if $b \geq 0$, and $\rho \to 8b^2/3$ if $b < 0$.
  Closed timelike curves are observed at large $x$ if $b =0$, $E_0 \ne 0$. A more detailed analysis is 
  beyond the scope of this paper.
      
\section{Conclusion}  

  This paper demonstrates the way of obtaining exact solutions for the gravitational fields of 
  rotating anisotropic fluids with pressures $p_i = w_i \rho$ ($i = 1,2,3$) under the assumption of 
  \cy\ symmetry. It is clear that this framework covers a diversity of physical systems such as 
  perfect fluids, cosmic string distributions, combinations of mutually opposite radiation flows 
  as well as some classical field  systems. The three examples briefly considered here show 
  a great diversity of physical properties of the solutions: some of them may describe \cy\ \whs, 
  others can possess a regular axis, and many of these space-times contain closed timelike
  curves. We postpone a detailed analysis of such solutions for the near future.      
    
\subsubsection*{Funding}

  This publication was supported by the RUDN University program 5-100 and by the 
  Russian Foundation for Basic Research Project 19-02-00346. 
  The work of KB was also performed within the framework of the Center FRPP 
  supported by MEPhI Academic Excellence Project (contract No. 02.a03.21.0005,
  27.08.2013).

\end{document}